\documentclass[showpacs,preprintnumbers,amsmath,amssymb]{revtex4}
\usepackage{graphicx}
\begin{document}

\title{Anomalous far infrared monochromatic transmission through a film of
type-II superconductor in magnetic field}
\author{Oleg L. Berman$^{1}$, Vladimir S. Boyko$^{1}$, Roman Ya. Kezerashvili%
$^{1}$, and Yurii E. Lozovik$^{2}$}
\affiliation{\mbox{$^{1}$Physics Department, New York City College
of Technology, the City University of New York,} \\
Brooklyn, NY 11201, USA \\
\mbox{$^{2}$ Institute of Spectroscopy, Russian Academy of
Sciences,} \\
142190 Troitsk, Moscow Region, Russia}

\begin{abstract}
Anomalous far infrared monochromatic transmission through a lattice
of Abrikosov vortices in a type-II superconducting film is found and
reported. The transmitted frequency corresponds to the photonic mode
localized by the defects of the Abrokosov lattice. These defects are
formed by extra vortices placed out of the nodes of the ideal
Abrokosov lattice. The extra vortices can be pinned by crystal
lattice defects of a superconductor. The corresponding frequency is
studied as a function of magnetic field and temperature in the
framework of the Dirac-type two-band model. While our approach is valid
for all type-II superconductors, the specific calculations have been
performed for the YBa$_{2}$Cu$_{3}$O$_{7-\delta}$ (YBCO). The control of
the transmitted frequency by varying magnetic field and/or
temperature is analyzed. It is suggested that found anomalously
transmitted localized mode can be utilized in the far infrared
monochromatic filters.
\end{abstract}

\maketitle



\section{Introduction}
\label{int}

Infrared spectroscopy is one of the most important analytical
techniques available to the modern science \cite{Stuart}. Due to
intensive development of infrared spectroscopy in the recent decade,
the construction of the novel types of far infrared monochromatic
filters attracts a strong interest. Extraordinary optical
transmission through nanostructures constructed as arrays of holes of
a subwavelength diameter in metal films has been the subject of
extensive study since detection of large enhancements in transmitted
intensity was first reported \cite{Ebbsen}. Several mechanisms
responsible for this enhancement have been discussed, including
excitation of surface plasmon polaritons (SPPs) of the film
surfaces \cite{Porto,Garcia}. Such nanostructures can be used as
monochromatic filters. However, it is impossible to control the
transmitted resonant frequencies in such devices by an external field.
In other words, these nanomaterials are not tunable.  In this Paper, we suggest an
idea of the new type of a tunable far infrared monochromatic
 filter consisting of extra vortices placed out of
the nodes of the ideal Abrikosov lattice. These extra vortices are
pinned by a crystal defects in a type-II superconductor in strong
magnetic field. The resonant transmitted frequencies can be
controlled by two ways: changing external magnetic field $B$ and
temperature $T$, because the critical magnetic field $B_{c2}$ depends parametrically on temperature.

Photonic crystals, artificial media with a spatially periodical
dielectric function that were first discussed by
Yablonovitch~\cite{Yablonovitch} and John~\cite{John}, are the
subjects of growing interest due to various modern applications
\cite{Eldada,Chigrin}. This periodicity can be achieved by embedding
a periodic array of constituent elements (``particles'') with
dielectric constant $\varepsilon_{1}$ in a background medium
characterized by dielectric constant $\varepsilon_{2}$. The first
experimental evidence of a photonic band structure was observed with
metallic meshes in the THz-range by Ulrich and Tacke \cite{Ulrich}.
Different materials have been used for the corresponding elements including
dielectrics, semiconductors and metals
\cite{Joannopoulos1,Joannopoulos2,Sun_Jung,Sun,Maradudin,Kuzmiak}.

Previous studies have investigated the photonic band gap structure
created by the propagation of light through a dielectric medium
characterized by some dielectric constant with periodically located
dielectric particles characterized by another dielectric
constant \cite{Joannopoulos1,Joannopoulos2}. The optical properties
of low-dimensional metallic structures have also been investigated
recently. For example, the optical transmission through a nanoslit
array structure formed on a metal layer with tapered film thickness
was analyzed in Refs.~[\onlinecite{Sun_Jung,Sun}]. The photonic band
structures of a square lattice array of metal or semiconductor
cylinders, and of an array of metal or semiconductor spheres, were
computed numerically in Ref.~[\onlinecite{Maradudin}]. However, a
photonic crystal formed by placing superconducting particles in the
nodes of the lattice has not been considered previously. Such a
system is interesting, particularlly, because of the unique optical
properties of superconductors (see, for example,
Refs.~[\onlinecite{Schrieffer,Abrikosov}]). In recent experiments
superconducting (SC) metals (in particular, Nb) have been used as
components in optical transmission nano-materials. It was found that
dielectric losses are substantially reduced in the SC metals
relative to analogous structures made out of normal metals. Also, it should be mentioned that
band edges tend to be sharper with the SC metals. The dielectric
losses of such  SC nano-material \cite{Ricci} were found to be
reduced by a factor of $6$ upon entering the SC state.

Photonic gaps are formed at frequencies $\omega$ at which the
dielectric contrast $\omega^{2}(\varepsilon_{1}(\omega) -
\varepsilon_{2}(\omega))$ is sufficiently large. Since the quantity
$\omega^{2}\varepsilon(\omega)$ is included into the electromagnetic wave
equation \cite{Joannopoulos1,Joannopoulos2}, only metal-containing
photonic crystals can maintain the necessary dielectric contrast at
small frequencies due to their Drude-like behavior
$\varepsilon_{Met} (\omega) \sim
-1/\omega^{2}$ \cite{Maradudin,Kuzmiak}.   However, the damping of
electromagnetic waves in metals can suppress many potentially useful
properties of metallic photonic crystals.

A novel type of photonic crystal consisting of  {\it
superconducting} elements embedded in a dielectric medium was
proposed in Ref.~[\onlinecite{Berman}]. Such photonic crystal
provides the photonic band gap tuned by an external magnetic field
and temperature. The photonic band spectrum of the  ideal triangular
Abrikosov lattices in type-II superconductors studied as photonic
crystals (ideal photonic crystal) has been calculated in
Ref.~[\onlinecite{Zakhidov}].

In this Paper, we calculate the photonic frequency spectrum of a
photonic crystal with an extra vortex out of the node of the
Abrikosov lattice (real photonic crystal) in a type-II
superconductor in a magnetic field. The problem is solved in two
steps: (1) we recall the procedure of the solution the eigenvalue
problem for the calculation of the photonic-band spectrum of the
ideal Abrikosov lattice \cite{Berman,Basov,Zakhidov}; (2) we apply
Kohn-Luttinger two-band model \cite{Luttinger,Kohn,Keldysh,Volkov}
to calculate the eigenfrequency spectrum of the Abrikosov lattice
with one extra vortex inside. Based on the results of our
calculations we are suggesting a new type of a tunable far infrared
monochromatic filter consisting of extra vortices placed out of the
nodes of the ideal Abrikosov lattice. These extra vortices are
pinned by a crystal defects in a type-II superconductor in strong
magnetic field. As a result of change of an external magnetic field
$B$ and temperature $T$  the resonant transmitted frequencies can be
controlled. This paper is organized as follows. In Sec.~\ref{ideal}
we analyze the mapping of the wave equation for the electromagnetic
wave penetrating through the ideal Abrikosov lattice onto the
Schr\"{o}dinger equation for the wavefunction of an electron in the
periodic field of a crystal lattice. In Sec.~\ref{real} we perform
the calculations of the eigenfrequency corresponding to the
electromagnetic wave localized due to the extra vortex  out of the
nodes of the ideal Abrokosov lattice applying Kohn-Luttinger
two-band model, and frequency for the anomalous far infrared monochromatic transmission is given.
 In Sec.~\ref{discussion} we discuss our results
proposing monochromatic filter based on type-II superconductor in
magnetic field. Conclusions follow in Sec.~\ref{conc}.


\section{The Abrikosov lattice as an ideal photonic crystal}
\label{ideal}

Let us consider a system of Abrikosov vortices in a type-II
superconductor that are arranged in a triangular lattice. We treat
Abrikosov vortices in a superconductor as the parallel cylinders of
the normal metal phase in the superconducting medium. The axes of
the vortices, which are directed along the $\hat z$ axis, are
perpendicular to the surface of the superconductor. We assume the
$\hat x$ and $\hat y$ axes to be parallel to the two real-space
lattice vectors that characterize the 2D triangular lattice of
Abrikosov vortices in the film and the angle between $\hat x$ and
$\hat y$ is equal $\pi/3$. The nodes of the 2D triangular lattice of
Abrikosov vortices are assumed to be situated on the $\hat x$ and
$\hat y$ axes.

\begin{figure}[tbp]
\includegraphics[width = 3.0in] {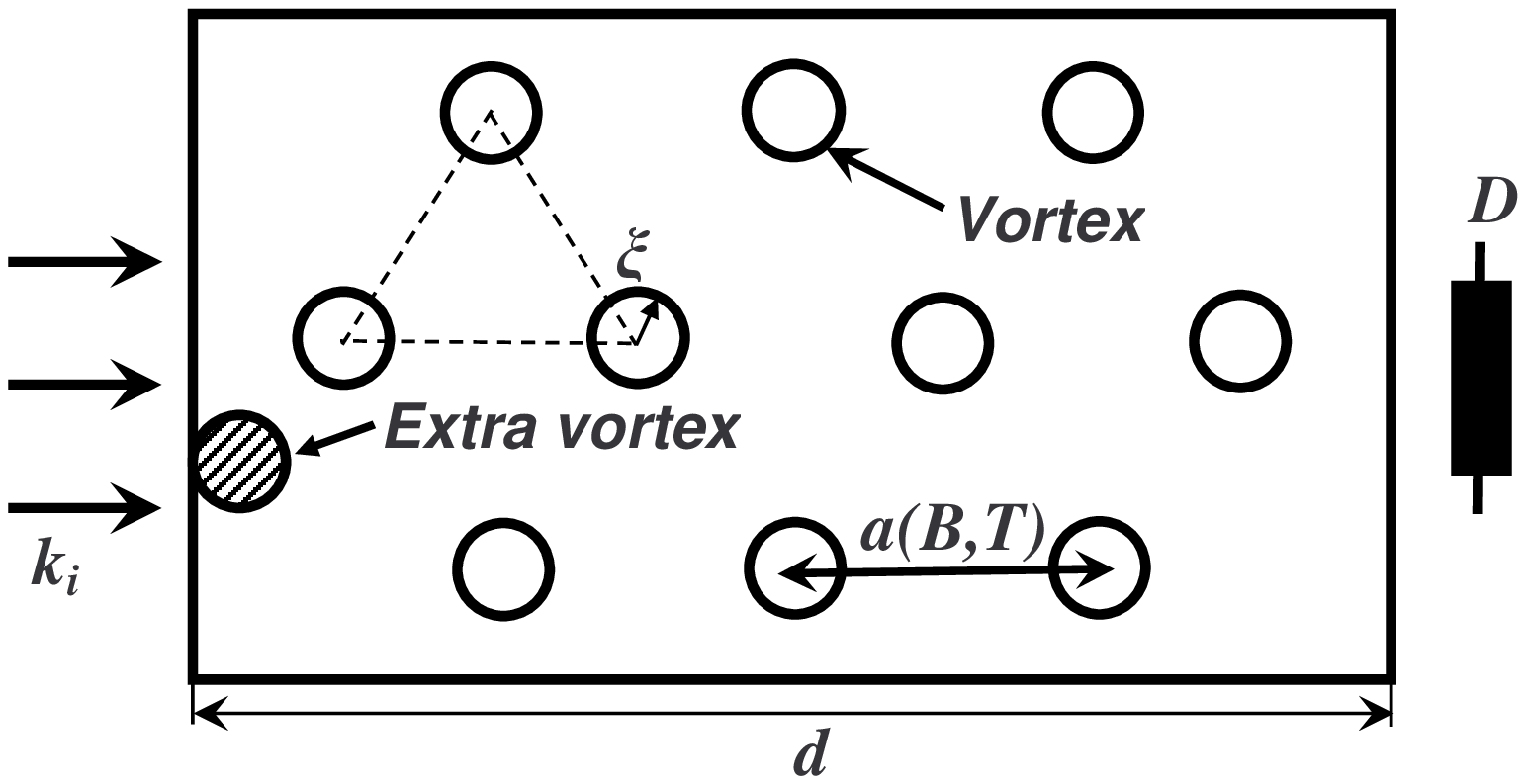}
\caption{Anomalous far infrared monochromatic transmission through a
film of type-II superconductor in the magnetic field parallel to the
vortices. $a (B,T)$ is the equilateral triangular Abrikosov lattice
spacing. $\protect\xi$ is the coherence length and the radius of the
vortex. $d$ denotes the length of the film. The shaded extra vortex placed
near the boundary of the film and situated outside of the node of the
lattice denotes the defect of the Abrikosov lattice.} \label{crystal}
\end{figure}

For simplicity, we consider the superconductor in the London
approximation \cite{Abrikosov} i.e. assuming that the London
penetration depth $\lambda $ of the bulk superconductor is much
greater than the coherence length $\xi $: $\lambda $ $\gg \xi$. Here
the London penetration depth is $\lambda =[m_{e}c^{2}/(4\pi
n_{e}e^{2})]^{1/2},$ where $n_{e}$ is electron density, $m_{e}$ and
$e$ are the mass and the charge of the electron, respectively.
The coherence length is defined as $\xi =c/(\omega _{p0}\sqrt{\epsilon })$, where $%
\omega _{p0}=2\pi c\omega _{0}$ is the plasma frequency, $c$ is the speed of
light.  A schematic diagram of Abrikosov lattices in type-II superconductors is shown in Fig.~\ref{crystal} (the presence of the pinned extra vortex is discussed in Sec.~\ref{real}). As it is seen
 from Fig.~\ref{crystal} the Abrikosov
vortices of radius $\xi $ arrange themselves into a 2D triangular lattice
with lattice spacing $a(B,T)=2\xi (T)\left[ \pi B_{c2}/(\sqrt{3}B)\right]
^{1/2}$ \cite{Zakhidov} at the fixed magnetic field $B$ and temperature $T$. Here $B_{c2}$ is the critical magnetic field for the superconductor.
We assume the wavevector of the incident electromagnetic wave vector $\mathbf{k}_{i}$ to be perpendicular to the direction of the Abrikosov
vortices and the transmitted wave can be detected by using the detector $D$.

Now let us follow the procedure used in Ref.~\cite{Berman} to obtain the wave equation for Abrikosov lattice treated as a two-component photonic crystal. In Ref.~[\onlinecite{Berman}], a system
consisting of superconducting cylinders in vacuum is studied. By contrast, the
system under study in the present manuscript consists of the cylindrical
vortices in a superconductor, which is a complementary case (inverse
structure) to what was treated in Ref.~[\onlinecite{Berman}].
For this system of the cylindrical vortices in the superconductor, we write the wave equation for the electric field $\mathbf{E}(x,y,t)$ parallel to the vortices in the form of 2D partial differential equation.
The corresponding wave equation for the electric field is
\begin{eqnarray}\label{subst}
-\nabla^{2}\mathbf{E} = - \frac{1}{c^2}\epsilon\sum_{\{%
\mathbf{n}^{(l)}\}}\eta (\mathbf{r})
\frac{\partial^{2}\mathbf{E}}{\partial t^{2}} -
\frac{4\pi}{c^{2}}\frac{\partial \mathbf{J}(\mathbf{r})}{\partial
t},
\end{eqnarray}
where $\epsilon$ is a dielectric constant of the normal metal component inside the vortices, $\eta (\mathbf{r})$ is the Heaviside step function
 which is $\eta (\mathbf{r})=1$ inside of the vortices and otherwise $\eta (\mathbf{r})=0$.  In Eq.~(\ref{subst}) $\mathbf{n}^{(l)}$ is a vector of integers that gives the
location of a scatterer $l$ at $\mathbf{a}(\mathbf{n}^{(l)})\equiv
\sum_{i=1}^{d}n_{i}^{(l)}\mathbf{a}_{i}$ ($\mathbf{a}_{i}$ are real
space lattice vectors situated in the nodes of the 2D triangular
lattice and $d$ is the dimension of Abrikosov lattice).

 At $(T_{c} -T)/T_{c} \ll 1$ and $\hbar \omega \ll \Delta
\ll T_{c}$, where $T_{c}$ is the critical temperature and $\Delta$ is the superconducting gap, a simple relation for the current density holds~\cite{Abrikosov}:
\begin{eqnarray}\label{sup_cur}
\mathbf{J}(\mathbf{r}) = \left[-\frac{c}{4\pi\delta_{L}^{2}} +
\frac{i\omega \sigma}{c}\right] \mathbf{A}(\mathbf{r}) .
\end{eqnarray}
In Eq.~(\ref{sup_cur}) $\sigma$ is the conductivity of the normal metal component.

The important property determining the band structure of the photonic crystal is the dielectric constant.
The dielectric constant, which depends on the frequency, inside and outside
of the vortex is considered in the framework of the two-fluid model. For a normal metal phase
inside of the vortex it is $\epsilon _{in}(\omega )$ and for a superconducting phase
outside of the vortex it is $\epsilon _{out}(\omega )$ and
can be described via a simple Drude model. Following Ref.~[\onlinecite{Zakhidov}] the dielectric constant can be written in the form:
\begin{eqnarray}\label{ein}
\epsilon _{in}(\omega )=\epsilon ,\hspace{1cm}\epsilon _{out}(\omega
)=\epsilon \left( 1-\frac{\omega _{p0}^{2}}{\omega ^{2}}\right) .
\end{eqnarray}
Eqs.~(\ref{ein}) are obtained in Ref.~[\onlinecite{Zakhidov}] from a phenomenological two-component fluid model \cite{Takeda2} by applying the
following condition: $\omega_{pn} \ll \omega \ll \gamma$. Here
$\omega_{pn}$ is the plasma frequency of normal conducting
electrons, and $\gamma$ is the damping term in the normal conducting
states.

Let's neglect a damping in the superconductor. After that
substituting Eq.~(\ref{sup_cur}) into Eq.~(\ref{subst}), considering
Eqs.~(\ref{ein}) for the dielectric constant, and seeking a solution
in the form  with harmonic time variation of the electric field,
i.e., $\mathbf{E}(\mathbf{r},t) =
\mathbf{E}_{0}(\mathbf{r})e^{i\omega t}$, $\mathbf{E} = i\omega
\mathbf{A}/c$, we finally obtain the following equation
\begin{equation}
\label{substparteq}
-\nabla ^{2}E_{z}(x,y)=\frac{\omega ^{2}\epsilon }{c^{2}}\left[ 1-\frac{%
\omega _{p0}^{2}}{\omega ^{2}}+\frac{\omega _{p0}^{2}}{\omega ^{2}}\sum_{\{%
\mathbf{n}^{(l)}\}}\eta (\mathbf{r})\right] E_{z}(x,y),
\end{equation}
where  $\omega $ is the frequency and $\omega_{p0}$ is the plasma frequency. The summation in Eq.~(\ref{substparteq}) goes over all
 lattice nodes characterizing positions of the Abrikosov vortices. Eq.~(\ref{substparteq}) describes Abrikosov lattice as the two-component 2D photonic crystal.
The first two terms within the bracket are associated to the
superconducting medium, while the last term is related to  vortices
(normal metal phase).
 Here and below the system described by Eq.~(\ref{substparteq}) will be defined as an ideal photonic crystal. The ideal photonic crystal based on the Abrikosov lattice
in type-II superconductor was studied in
Refs.~[\onlinecite{Zakhidov,Takeda2,Takeda1}]. The wave
equation~(\ref{substparteq}) describing the Abrikosov lattice has
been solved in Ref.~[\onlinecite{Zakhidov}] where the photonic band
frequency spectrum $\omega =\omega (\mathbf{k})$ of the ideal
photonic crystal of the vortices has been calculated.

The wave equation~(\ref{substparteq}) for the electric field can be mapped
onto the 2D Schr\"{o}dinger equation for an electron with effective mass $m_{0}$ in the periodic potential  $W(\mathbf{r})$ of the 2D crystal lattice:
\begin{eqnarray}
\label{sch}
\left[ -\frac{\hbar ^{2}}{2m_{0}}\nabla ^{2}+W(\mathbf{r})\right] \psi_{0}(x,y)= \varepsilon _{\omega }\psi _{0}(x,y).
\end{eqnarray}
As a result of mapping in Eq. (\ref{sch}) the eigenfunction $\psi _{0}(x,y)=E_{z}(x,y)$, the periodic potential $W(\mathbf{r})$ is
\begin{eqnarray}\label{perpot}
W(\mathbf{r})=-\hbar ^{2}\epsilon \omega _{p0}^{2}/(2c^{2}m_{0})\sum_{\{\mathbf{n}%
^{(l)}\}}\eta (\mathbf{r})
\end{eqnarray}
and the eigenenergy  is
\begin{eqnarray}\label{eigenen}
\varepsilon _{\omega}=\hbar ^{2}\epsilon \left[ \omega ^{2}-\omega _{p0}^{2}\right]
(2m_{0}c^{2})^{-1} \ .
\end{eqnarray}
It is important to note that according to Eq.~(\ref{eigenen}) the eigenenergy of such electron $\varepsilon
_{\omega }$ depends on the frequency $\omega $.

Let us expand the periodic wave function
 $\psi_{0} (\mathbf{r})$ in terms of $ u_{n0} (\mathbf{r})$, which are the periodic solutions of Eq. (\ref{sch}) corresponding to $k = 0$
\begin{eqnarray}  \label{psi0}
\psi_{0} (\mathbf{r}) = \sum_{n\mathbf{k}} c_{n}(\mathbf{k}) \exp\left[i%
\mathbf{k}\mathbf{r}/\hbar \right] u_{n0} (\mathbf{r}) \ ,
\end{eqnarray}
where $c_{n}(\mathbf{k})$ are the coefficients of the expansion, which can be determined as a result of substitution Eq.~(\ref{psi0}) into Eq. (\ref{sch}), and $n$ indicates the number of the band.


\section{The Abrikosov lattice with an extra vortex as a real photonic crystal}
\label{real}

Let us consider an extra Abrikosov vortex pinned by some defect in the type-II superconducting
material, as shown in Fig.~\ref{crystal}. This extra vortex contributes to the dielectric contrast by the
adding the term $\epsilon \omega _{p0}^{2}/c^{2}\eta (\xi -|\mathbf{r}-%
\mathbf{r}_{0}|)E_{z}(x,y)$, where $\mathbf{r}_{0}$ points out the position of the extra vortex, to the r.h.s. in Eq.~(\ref{substparteq}):
\begin{equation}
\label{realcrys}
-\nabla ^{2}E_{z}(x,y)=\frac{\omega ^{2}\epsilon }{c^{2}}\left[ 1-\frac{%
\omega _{p0}^{2}}{\omega ^{2}}+\frac{\omega _{p0}^{2}}{\omega ^{2}}\sum_{\{%
\mathbf{n}^{(l)}\}}\eta (\mathbf{r}) +\frac{\omega _{p0}^{2}}{\omega ^{2}}\eta (\xi -|\mathbf{r}-%
\mathbf{r}_{0}|) \right] E_{z}(x,y),
\end{equation}
Eq.~(\ref{realcrys}) describes the type-II superconducting medium with the extra Abrikosov vortex pinned by a defect in the superconductor.  We define the photonic crystal implying an extra Abrikosov vortex pinned by a defect as a real photonic crystal and it is  described by Eq.~(\ref{realcrys}).

Let's mention that the addition of the extra vortex pinned by some defect leads to a modification of the dielectric constant, as it follows from Eq.~(\ref{realcrys}). However, in our consideration the defect pinning the
extra vortex does not effect on the dielectric constant of the
normal and superconducting components. Besides, let us emphasize that we consider no external current in the system.

   After mapping of Eq.~(\ref{realcrys}) onto the Schr\"{o}dinger equation
   for an electron with the effective electron mass $m_{0}$ we have
\begin{eqnarray}\label{sch1}
\left[ -\frac{\hbar ^{2}}{2m_{0}}\nabla ^{2}+W(\mathbf{r})+V(\mathbf{r})%
\right] \psi (x,y)=\varepsilon _{\omega }\psi (x,y).
\end{eqnarray}
In Eq.(\ref{sch1}) $\psi (x,y)=E_{z}(x,y)$, and  the potential $V(\mathbf{r})$ is defined as
\begin{eqnarray}\label{imppot}
V(\mathbf{r})=- V_{0} \eta (\xi -|\mathbf{r}-\mathbf{r}_{0}|) \ , \ \ \ \ \ \ \ \ \
 V_{0} = \hbar ^{2}\epsilon \omega
_{p0}^{2}/(2m_{0}c^{2}).
\end{eqnarray}
Eq.~(\ref{sch1}) has the same form as Eq.~(6) in
Ref.~[\onlinecite{Keldysh}]. However, in our case the potential
$V(\mathbf{r})$ is defined by Eq.~(\ref{imppot}) and corresponds  to
the potential of the impurity in the Schr\"{o}dinger equation for an
\textquotedblleft electron\textquotedblright\ in the periodic field
of the crystal lattice and in the presence of the ``impurity''.

Since the contribution to the dielectric contrast $\epsilon \omega
_{p0}^{2}/c^{2}$ from an extra Abrikosov vortex has the same order
of magnitude as the photonic band gap $\tilde{\Delta}$ of the ideal
Abrikosov lattice  calculated in Ref.~[\onlinecite{Zakhidov}], we
expect the eigenfrequency level corresponding to the extra vortex to
be situated inside the photonic band gap. Our calculations will
demonstrate below that this expectation holds.

Let us apply to Eq.~(\ref{sch1}) the two-band model \cite{Keldysh}, where  two different neighboring
photonic bands are described by wave functions $\varphi (\mathbf{r})$ and $\chi (\mathbf{r})$
and, therefore, introduce two-components spinor as
\begin{eqnarray}
\label{2comp}
 \psi (\mathbf{r})=\left(
\begin{array}{c}
\varphi (\mathbf{r}) \\
\chi (\mathbf{r})%
\end{array}%
\right) \ .
\end{eqnarray}
Note that Eq.~(\ref{sch1}) describes an electron in the periodic potential of the ideal crystal
 lattice $W(\mathbf{r})$ and the potential of the impurity $V(\mathbf{r})$.
 If the solution corresponding to the absence of impurity $V(\mathbf{r}) = 0$ is known,
 the energy levels of the electron localized by the ``impurity'' can be obtained by replacing Eq.~(\ref{sch1})
 by the Dirac-type equation according to Luttinger-Kohn model described in Refs.~[\onlinecite{Luttinger,Kohn,Keldysh}].
  This model implies Direc-type equation for the two-component spinor wave
  function. According to Ref.~\cite{Keldysh}, the function
  $\varphi_{n}(\mathbf{r})$ defined as
\begin{eqnarray}
\label{phid} \varphi_{n}(\mathbf{r}) = \sum_{\mathbf{k}} c_{n}(\mathbf{k}) \exp\left[i%
\mathbf{k}\mathbf{r}/\hbar \right]
\end{eqnarray}
satisfies to the set of the second order partial differential equations. Considering only two neighboring bands corresponding to  the wavefunction $\psi (\mathbf{r})$ given by Eq.~(\ref{2comp}) and described by wavefunctions $\varphi (\mathbf{r})$ and $\chi (\mathbf{r})$ in the limit
$|\varepsilon_{\omega}^{2} -
\Delta_{\omega}^{2}|/(2\Delta_{\omega}^{2}) \ll 1$, this set of
equations for  $\varphi_{n}(\mathbf{r})$ can be reduced to the
Dirac-type equations for the two-component spinor~(\ref{2comp}), which has the
following form \cite{Keldysh}
\begin{eqnarray}
\left[ \varepsilon _{\omega }-\Delta _{\omega }-V(\mathbf{r})\right]
\varphi (\mathbf{r})+i\hbar s\boldsymbol{\sigma} \cdot
\boldsymbol{\nabla }\chi (\mathbf{r}) &=&0,
\nonumber  \label{dir} \\
\left[ \varepsilon _{\omega }+\Delta _{\omega }-V(\mathbf{r})\right] \chi (%
\mathbf{r})+i\hbar s\boldsymbol{\sigma} \cdot \boldsymbol{\nabla
}\varphi (\mathbf{r}) &=&0,
\end{eqnarray}
In Eqs.~(\ref{dir}), as it follows from the mapping of the wave equation for the electric field in the Abrikosov lattice onto Eq.~(\ref{sch1}),
\begin{eqnarray}
\label{deltaomeg}
\Delta _{\omega }=\hbar ^{2}\epsilon \lbrack \tilde{\Delta}%
^{2}-\omega _{p0}^{2}]/(2m_{0}c^{2})
\end{eqnarray}
 is the forbidden band in the
electron spectrum defined by Eq.~(\ref{eigenen}), and
$\boldsymbol{\sigma }$ are the Pauli matrices defined as
\begin{eqnarray}
\label{pauli}
\sigma _{x}=\left(
\begin{array}{cc}
0 & 1\  \\
1 & 0\
\end{array}%
\right) ,\hspace{1cm}\sigma _{y}=\left(
\begin{array}{cc}
0 & -i\  \\
i & 0\
\end{array}%
\right) \ .
\end{eqnarray}
In Eq.~(\ref{dir}) $s=p/\left( \sqrt{3}m_{0}\right) $ has the
dimension of the velocity, and $\mathbf{p}$ is given by
$(p_{cv\alpha })_{\beta } = \hbar k_{0}\delta _{\alpha \beta }$,
where
\begin{eqnarray}
\label{pm}
k_{0}\delta _{\alpha \beta } &=&-i\left( \left[ \int u_{c0}^{\ast }(\mathbf{r%
})\boldsymbol{\nabla }u_{v0}(\mathbf{r})d^{2}r\right] _{cv\alpha }\right)
_{\beta } \ ,
\end{eqnarray}%
and $\alpha ,\beta $ can be $x$ or $y$. The index $c$ corresponds to
the upper photonic band, and the index $v$ corresponds to the lower
photonic band. Let us mention that the two-component model
\cite{Keldysh} described by the Dirac-type equations (\ref{dir}) is
necessary only for the ``deep impurities'', when the potential of the
impurity has the same order of magnitude as the forbidden band:
$|V_{0}/\Delta _{\omega }| \sim 1$. Note that in the limit, when the
potential of the impurity would be much smaller than the gap,  these
Dirac-type equations would be reduced to the effective
Schr\"{o}dinger equation for the scalar wave function corresponding
to the effective mass approximation.

We have reduced the problem of the Schr\"{o}dinger equation for a
particle in the periodic potential  $W(\mathbf{r})$  related to the system of the
periodically placed vortices and an ``impurity potential''
$V(\mathbf{r})$ related to an extra vortex to much more simple
equation for the envelope wavefunctions. These wavefunctions imply
the existence of two bands and contain only the potential of an
impurity $V(\mathbf{r})$, while the periodic potential
$W(\mathbf{r})$ enters only in the effective velocity $s$. Taking
into account the two-band structure, the equation for two-component
spinor wavefunction $\psi (\mathbf{r})$ has the form provided by
Eq.~(\ref{2comp}).  Note that Eq.~(\ref{sch1}) contains both the
periodic function $W(\mathbf{r})$ corresponding to the ideal lattice
and $V(\mathbf{r})$, describing the potential of an ``impurity''.
Without an impurity the energy spectrum would be described by two
neighboring bands and the gap between them. Taking into account an
impurity, we have reduced the problem to the approximation
generalizing the effective mass approximation and implying the two
band structure. Applying the standard two-band approach, we have
obtained an effective Dirac-type equation (\ref{dir}) for the
envelope spinor wavefunction, which imply the periodicity provided
by $W(\mathbf{r})$.

The condition for ``deep impurities'' is valid for an extra Abrikosov
vortex only if $\omega_{up}^{2}/\tilde{\Delta}_{\omega}^{2} \sim 1
$, which is true \cite{Zakhidov}. Note that $\omega _{up}(x)$ and
$\omega _{dn}(x)=\omega _{up}(x)-\tilde{\Delta}(x)$ are the up and
down boundaries of the photonic band gap, correspondingly
\cite{Zakhidov}.

Defining the effective mass of a quasiparticle as
\begin{eqnarray}
\label{momeg}
m_{\omega }=\Delta
_{\omega }/s^{2}=3m_{0}^{2}\Delta _{\omega }/p^{2} \ ,
\end{eqnarray}
 applying $i\hbar
\partial /\partial t$ to the l.h.s. and the r.h.s. of Eq.~(\ref{pauli}), and
following   the standard procedure of the quantum electrodynamics
\cite{Fermi,Bjorken} we obtain from the system of Dirac-type  Eqs.~(%
\ref{pauli}) the following Klein-Gordon type equation
\begin{eqnarray}
\label{kge}
\left[ -\hbar ^{2}s^{2}\nabla ^{2}+2m_{\omega }s^{2}V(\mathbf{r})\right]
\Psi (\mathbf{r})=(\varepsilon _{\omega }^{2}-m_{\omega }^{2}s^{4})\Psi (%
\mathbf{r}) \ ,
\end{eqnarray}
This Klein-Gordon type equation has the form of the 2D
Schr\"{o}dinger equation for a particle in the cylindrical potential
well with the eigenvalue
\begin{eqnarray}
\label{eomeg}
\mathcal{E}_{\omega }=(\varepsilon
_{\omega }^{2}-m_{\omega }^{2}s^{4})/(2m_{\omega }s^{2}) \ .
\end{eqnarray}
 The set of the eigenvalues $\mathcal{E}_{\omega }^{(nm)}$ and eigenfunctions
$\Psi^{(nm)}$ correspond to the quantum numbers $n=0,1,2,\ldots$, $m = \ldots, -2,-1,0,1,2, \ldots$. Our particular interest is only discrete eigenstate corresponding to the localized eigenfunction, which is
characterized by the lowest discrete eigenvalue $\mathcal{E}_{\omega
}^{(00)}$. Eq.~(\ref{kge}) was solved
in Ref.~[\onlinecite{Flugge}] and the solution for the discrete lowest
eigenstate is
\begin{eqnarray}\label{eigen_v}
\mathcal{E}_{\omega }^{(00)}=-\frac{2\hbar ^{2}}{m_{\omega }\xi ^{2}}\exp
\left( -\frac{2\hbar ^{2}}{m_{\omega }\xi ^{2}V_{0}}\right) ,
\end{eqnarray}
and
\begin{eqnarray}\label{eigen_f}
&&\Psi^{(00)}(\mathbf{r})=   \\
&&\left\{
\begin{array}{l}
C_{1},\ |\mathbf{r}-\mathbf{r}_{0}|<\xi \ , \\
C_{2}\log \left[ 2\hbar \left( 2m_{\omega }|\mathcal{E}_{\omega
}^{(00)}|\right) ^{-1/2}|\mathbf{r}-\mathbf{r}_{0}|^{-1}\right] ,\ |\mathbf{r%
}-\mathbf{r}_{0}|>\xi \ ,%
\end{array}%
\right.   \nonumber
\end{eqnarray}
where the constants $C_{1}$ and $C_{2}$ can be obtained from the condition
of the continuity of the function $\Psi ^{(00)}(r)$ and its derivative at
the point $|\mathbf{r}-\mathbf{r}_{0}|=\xi $.

In terms of the initial quantities of the Abrikosov lattice the
eigenfrequency $\omega $ of the localized photonic state can be
obtained by substituting Eqs.~(\ref{deltaomeg}),~(\ref{momeg})
and~(\ref{eomeg}) into Eq.~(\ref{eigen_v}). As the result, we
finally obtain
\begin{eqnarray}\label{eigen_fr}
\omega (x)=\left( \omega _{up}^{4}(x)-A(x)\right) ^{1/4},
\end{eqnarray}
where $x=B/B_{c2}$ and function $A(x)$ is given by
\begin{eqnarray}\label{eigen_fr_A}
A(x)=\frac{16c^{4}k_{0}^{2}(x)}{3\epsilon ^{2}\xi ^{2}}\exp \left[ -\frac{%
8k_{0}^{2}(x)c^{4}}{3\epsilon ^{2}\omega _{up}^{2}(x)\xi ^{2}\omega _{p0}^{2}%
}\right]
\end{eqnarray}
and  $k_{0}(x)$ in Eq.~(\ref{eigen_fr_A}) can be obtained from Eq. (\ref{pm}) and is defined below
through the electric field of the lower and higher photonic bands of the
ideal Abrikosov lattice.

The electric field $E_{z}(x,y)$ corresponding to this localized photonic
mode can be obtained by substituting the initial quantities of the Abrikosov
lattice from Eqs.~(\ref{deltaomeg}),~(\ref{momeg}) and~(\ref{eomeg}) into Eq.~(\ref{eigen_f}), and we get:
\begin{eqnarray}
\label{eigen_f22} && E_{z}^{(00)}(\mathbf{r}) =  \left\{
\begin{array}{cc}
\tilde{C}_{1}, & |\mathbf{r}-\mathbf{r}_{0}|<\xi \ , \\
\tilde{C}_{2}B(x), & |\mathbf{r}-\mathbf{r}_{0}|>\xi \ ,%
\end{array}
\right.
\end{eqnarray}
where the constants $\tilde{C}_{1}$ and $\tilde{C}_{2}$ can be obtained from
the condition of the continuity of the function $E_{z}^{(00)}(\mathbf{r})$
and its derivative at the point $|\mathbf{r}-\mathbf{r}_{0}|=\xi $ and
\begin{eqnarray}
&&B(x)=\log \left[ \xi \left( |\mathbf{r}-\mathbf{r}_{0}| \times
\right. \right.
\label{eigen_f223} \\
&&\left. \left. \exp \left[ -4k_{0}^{2}(x)c^{4}/\left( 3\epsilon ^{2}\left(
\tilde{\Delta}^{2}(x)-\omega _{p0}^{2}\right) \xi ^{2}\omega
_{p0}^{2}\right) \right] \right) ^{-1}\right] .  \nonumber
\end{eqnarray}%
The function $k_{0}(x)$ is given as (see Eq.(\ref{pm}))
\begin{eqnarray}
k_{0}\delta _{\alpha \beta }=-i\left( \left[ \int E_{zc0}^{\ast }(\mathbf{r}%
)\nabla E_{zv0}(\mathbf{r})d^{2}r\right] _{cv\alpha }\right) _{\beta },
\end{eqnarray}%
where $E_{zc0}(\mathbf{r})$ and $E_{zv0}(\mathbf{r})$ are defined by the
electric field of the up and down photonic bands of the ideal Abrikosov
lattice. The exact value of $k_{0}$ can be calculated by
substituting the electric field $E_{zc0}(\mathbf{r})$ and $E_{zv0}(\mathbf{r}%
)$ from Ref.~[\onlinecite{Zakhidov}]. Applying the weak coupling
model \cite{Abrikosov} corresponding to the weak dielectric contrast
between the vortices and the superconductive media $\omega
_{p0}^{2}/\omega
^{2}\left\vert \sum_{\{\mathbf{n}^{(l)}\}}\eta (\mathbf{r})-1\right\vert \ll 1$ we use the approximate estimation of $k_{0}$ in our calculations as $%
k_{0}(x)\approx 2\pi /a(x)=\pi \xi ^{-1}\sqrt{\sqrt{3}x/\pi }$.

Note that according to Eqs.~(\ref{eigen_f22})
and~(\ref{eigen_f223}), the maximum of the  electric field
corresponding to this localized mode is located at the center of the
extra Abrikosov vortex pinned by a defect. This electric field
$E_{z}^{(00)}(\mathbf{r})$ increases as applied magnetic field $B$
increases.


\section{Results and Discussion}
\label{discussion}

\begin{figure}[tbp]
\includegraphics[width = 3.0in] {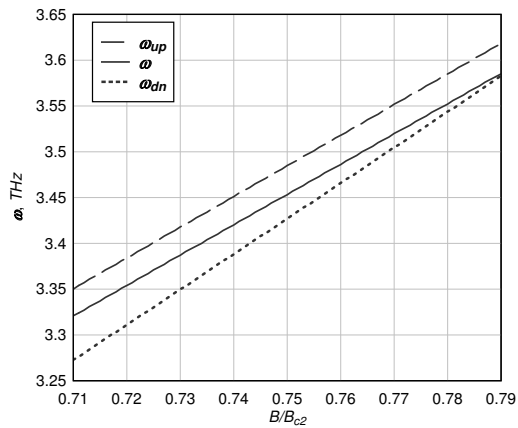}
\caption{The dependence of the photonic band structure of the real Abrikosov lattice
on $B/B_{c2}$. Solid line represents the eigenfrequency $\protect%
\omega$ corresponding to the localized mode near the extra vortex in the real
Abrikosov lattice given by Eq.~(\protect\ref{eigen_fr}). The dashed and dotted lines represent, respectively,
 the top $\protect\omega_{up}$ and bottom $\protect\omega_{dn}$ boundaries of the photonic
band gap of the ideal Abrikosov lattice according to Ref.~[
\onlinecite{Zakhidov}]. }
\label{omega}
\end{figure}

The approach developed in Sec.~\ref{real} we apply to the YBa$_{2}$Cu$_{3}$O$_{7-\delta}$ (YBCO) and study the dependence of the photonic band structure on the magnetic field. For the YBCO  the characteristic critical magnetic field $B_{c2}=5 \ \mathrm{T}$ at
temperature $T=85 \ \mathrm{K}$ is determined experimentally in Ref.~\cite{Safar}. So we obtained the frequency corresponding to the localized wave for the
YBCO in the magnetic field range from $B=0.72B_{c2}=3.6 \ \mathrm{T}$ up to $%
B=0.85B_{c2}=4.25\ \mathrm{T}$ at $T=85 \ \mathrm{K}$. Following
Ref. \cite{Zakhidov}, in our calculations we use the estimation
$\epsilon =10$ inside the vortices and for the YBCO $\omega_{0}/c=77
\ \mathrm{cm}^{-1}$. The dielectric contrast between the normal
phase in the core of the Abrikosov vortex and the superconducting
phase given by Eq.~(\ref{ein}) is valid only for the frequencies
below $\omega _{c1}$: $\omega <\omega _{c1}$, where $\omega _{c1}
=2\Delta _{S}/(2\pi \hbar )$, $\Delta _{S}=1.76k_{B}T_{c}$ is the
superconducting gap, $k_{B}$ is the Boltzmann constant, and $T_{c}$
is the critical temperature. For the YBCO we have $T_{c}=90 \
\mathrm{K}$, and $\omega <\omega _{c1}=6.601 \ \mathrm{THz}$. It can
be seen from Eqs.~(\ref{eigen_fr}) and~(\ref{eigen_f22}), that there
is a photonic state localized on the extra Abrikosov vortex, since
the discrete eigenfrequency corresponds to the electric field
decreasing as logarithm of the distance from an extra vortex. This
logarithmical behavior of the electric field follows from the fact
that it comes from the solution of 2D Dirac equation. The
calculations of   the eigenfrequency $\omega$ dependence on the
ratio $B/B_{c2}$, where $B_{c2}$ is the critical magnetic field, is
presented in Fig.~\ref{omega}. According to Fig.~\ref{omega}, our
expectation that the the eigenfrequency level $\omega$ corresponding
to the extra vortex is situated inside the photonic band gap is
true. We calculated the frequency corresponding to the localized
mode, which satisfies to the condition of the validity of the
dielectric contrast given by Eq.~(\ref{ein}). According to
Eqs.~(\ref{eigen_f22}) and~(\ref{eigen_f223}), the localized field
is decreasing proportionally to $\log
|\mathbf{r}-\mathbf{r}_{0}|^{-1}$ as the distance from an extra
vortex increases. Therefore, in order to detect this localized mode,
the length of the film $d$ should not exceed approximately
$10a(B/B_{c2})$, which corresponds to $d\lesssim 200 \ \mathrm{\mu
m}$ for the range of magnetic fields for the YBCO presented by
Fig.~\ref{omega}. For these magnetic fields $a\approx 20 \
\mathrm{\mu m}$. Since the frequency corresponding to the localized
mode is situated inside the photonic band gap, the extra vortex
should be placed near the surface of the film as shown in
Fig.~\ref{crystal}. Otherwise, the electromagnetic wave cannot reach
this extra vortex.  Besides, we assume that this localized photonic
state is situated outside the one-dimensional band of the surface
states of two-dimensional photonic crystal. It should be mentioned,
that in a case of several extra vortices separated at the distance
greater than the size of one vortex (it is the coherence length
estimated for the YBCO by $\xi \approx 6.5 \ \mathrm{\mu m}$) the
localized mode frequency for the both vortices is also going to be
determined by Eq.~(\ref{eigen_fr}). The intensity of the localized
mode in the latter case is going to be enhanced due to the
superposition of the modes localized by the different vortices. In
the case of far separated extra vortices we have neglected by the
vortex-vortex interaction. Thus, the existence of other pinned by
crystal defects vortices increases the intensity of the transmitted
mode and improves the possibility of this signal detection. Note
that at the frequencies $\omega$ inside the photonic band gap
$\omega_{dn}< \omega <\omega_{up}$
 the transmittance and reflectance of electromagnetic
waves would be close to zero and one, correspondingly, everywhere
except the resonant frequency $\omega$ related to an extra vortex.
The calculation of the transmittance and reflectance of
electromagnetic waves at this resonant frequency $\omega$ is a very
interesting problem, which will be analyzed elsewhere.

Let us mention that as $B/B_{c2}$ increases,  $\omega$
asymptotically converges to $\omega _{dn}$ up to the crossing point
of $\omega _{up}$ and $\omega _{dn}$ at $B/B_{c2} \approx 0.85$
\cite{Zakhidov}. The reason for this convergence of $\omega$ and
$\omega _{up}$ with the increment of  $B/B_{c2}$ is that  the
symmetric potential always implies the discrete level corresponding
to the eigenenergy of a particle in a 2D space \cite{Landau}.

The extra Abrikosov vortex resulting in the defect in photonic
crystal can be pinned by crystal lattice defect, for example,
dislocation \cite{Henry}. It is shown that the presence of the extra
vortex qualitatively influences the optical properties of the
type-II superconductor in external magnetic field. Hence, it would
be very useful to analyze the influence of different dislocation
microstructures on the specific optical transmission in analyzed
materials. The principles of a microdesign of twinning dislocation
structures for superconductive properties improvement of the YBCO in
magnetic fields have been analyzed in Ref.~[\onlinecite{Boyko}].
Magnetooptics in near fields as well as a neutron scattering \cite{Gilardi} seem to be useful for detecting the
extra vortex.

Above we presented the calculations for the YBCO. However,
it should be mentioned that our approach can be applied to a wide variety of
 type-II superconducting  materials, where Abrikosov lattice exists in the range of magnetic field $B_{c1}<B<B_{c2}$.
While Fig.~\ref{omega} implies the plasma frequency $\omega_{p0}$
corresponding to YBCO, the behavior shown in Fig.~\ref{omega} is
general and valid for any type-II superconducting films.
Just in this case we should replace the YBCO plasma frequency  $\omega_{p0}$ by the plasma frequency corresponding to the other type-II superconductor.

\section{Conclusions}
\label{conc}

We considered a type-II superconducting medium with an extra
Abricosov vortex pinned by a defect in a superconductor. By applying
mapping of the corresponding electromagnetic wave equation onto the
two-band model, the Dirac type equation was obtained. Solving this
equation we theoretically demonstrated the properties of such
Abrikosov lattices as real photonic crystals. The discrete photonic
eigenfrequency corresponding to the localized photonic mode, is
calculated as a function of the ratio $B/B_{c2}$, which
parametrically depends on temperature. This photonic frequency
increases as the ratio $B/B_{c2}$ and temperature $T$ increase.
Moreover, since the localized field and the corresponding photonic
eigenfrequency depend on the distance between the nearest Abrikosov
vortices $a(B,T)$, the resonant properties of the system can be
tuned by control of the external magnetic field $B$ and temperature
$T$. Based on the results of our calculations we can conclude that
it is possible to obtain a new type of a tunable far infrared
monochromatic filter consisting of extra vortices placed out of the
nodes of the ideal Abrikosov lattice, which can be considered as
real photonic crystals. These extra vortices are pinned by a crystal
defects in a type-II superconductor in strong magnetic field. As a
result of change of an external magnetic field $B$ and temperature
$T$  the resonant transmitted frequencies can be controlled.


\section*{Acknowledgements}
 Yu. E.~L. has been supported by the RFBR grants.


\end{document}